\documentclass{sbrt2019eng}

\usepackage[table,xcdraw]{xcolor}
\usepackage[tight, footnotesize]{subfigure}

\begin{document}
\title{Virtualized C-RAN Orchestration with Docker, Kubernetes and OpenAirInterface}

\author{Camila Novaes, Cleverson Nahum, Igor Trindade, Daniel Cederholm, Gyanesh Patra
and Aldebaro Klautau
\thanks{This work was supported in part by the Innovation Center, Ericsson
Telecomunica\c{c}\~oes S.A., Brazil, CNPq and the Capes Foundation, Ministry of Education of
Brazil, and by the European Union through the 5G-Crosshaul project (H2020-ICT-2014/671598).
C. Novaes, C. Nahum, I. Trindade and A. Klautau are with LASSE - 5G Group, Federal University
of Para, Belem, Brazil (e-mails: \{camila.novaes.silva, igor.trindade\}@itec.ufpa.br
\{cleversonahum, aldebaro\}@ufpa.br). D. Cederholm and G. Patra are with Ericsson Research,
Kista, Sweden (e-mails:\{daniel.cederholm, gyanesh.patra\}@ericsson.com).}}

\maketitle

\markboth{XXXVII SIMP\'OSIO BRASILEIRO DE TELECOMUNICA\c{C}\~OES E PROCESSAMENTO DE SINAIS -
SBrT2019, 29/09/2019--02/10/2019, PETR\'OPOLIS, RJ}{XXXVII SIMP\'OSIO BRASILEIRO DE
TELECOMUNICA\c{C}\~OES E PROCESSAMENTO DE SINAIS - SBrT2019, 29/09/2019--02/10/2019,
PETR\'OPOLIS, RJ}

\begin{abstract}
Virtualization is a key feature in Cloud Radio Access Network (C-RAN). It can help to save costs
since it allows the use of virtualized base stations instead of physically deployment in
different areas. However, the creation and management of virtual base stations pools is not
trivial and introduces new challenges to C-RAN deployment. This paper reports a method to
orchestrate and manage a container-based C-RAN. We used several instances of the
OpenAirInterface software running on Docker containers, and orchestrated them using Kubernetes.
We demonstrate that using Kubernetes it is possible to dynamically scale remote radio heads
(RRHs) and baseband units (BBUs) according to requirements of the network and parameters such
as the server resources usage.
\end{abstract}

\section{Introduction}

Cloud Radio Access Network (C-RAN)~\cite{checko_cloud_2015} are a new design for the cellular
architecture. It is an evolution of the distributed base stations concept in which baseband
and network processing are moved to a centralized processing center, creating a virtual base
station pool~\cite{checko_cloud_2015}. Therefore, the original base station (BS) is split into
centralized baseband units (BBU) with all baseband processing functions and remote radio head
(RRH) at the network edge, which contains radio functions, and the communication between these
two units is made through the fronthaul (FH) link.

In the virtual base station pool, an attractive feature is the possibility of sharing the
processing resources among different base stations, which may help to decrease costs. However,
by virtualizing the BS we introduce the need to properly manage this new infrastructure.
Therefore, we present in this paper a method to orchestrate a C-RAN infrastructure by applying
containerization on OpenAirInterface (OAI)~\cite{openairinterface} and using
Kubernetes~\cite{kubernetes}, an existing container orchestration platform.


\section{OAI, Virtualization and Orchestration}
The OAI is an open-source project created by EURECOM that aims to implement the $4^{th}$ and
$5^{th}$ generation of mobile cellular systems. It is fully compliant with the 3GPP LTE
standards. Thus, this platform can be used for 4G/5G research and development of new and
advanced features, as it provides a complete software implementation of all elements of the
4G system architecture: user equipment (UE), eNodeB (eNB), and the Core Network (EPC).

OAI is divided into two projects: \textit{openair-cn} and \textit{openairinterface5G}.
The first implements the Core Network and the second implements the Radio Access Network
(RAN) functionality. OAI supports connectivity to actual UEs (smartphones), using a commercial
software-defined radios (e.\,g. USRP). Alternatively, the OpenAirInterface System Emulation
(OAISIM) module can be used to emulate a UE. OAISIM is useful for scenarios when one needs
to change the UE software to test new features~\cite{batista201970}.

\begin{figure*}[htb]
\begin{center}
    \includegraphics[width=0.65\textwidth]{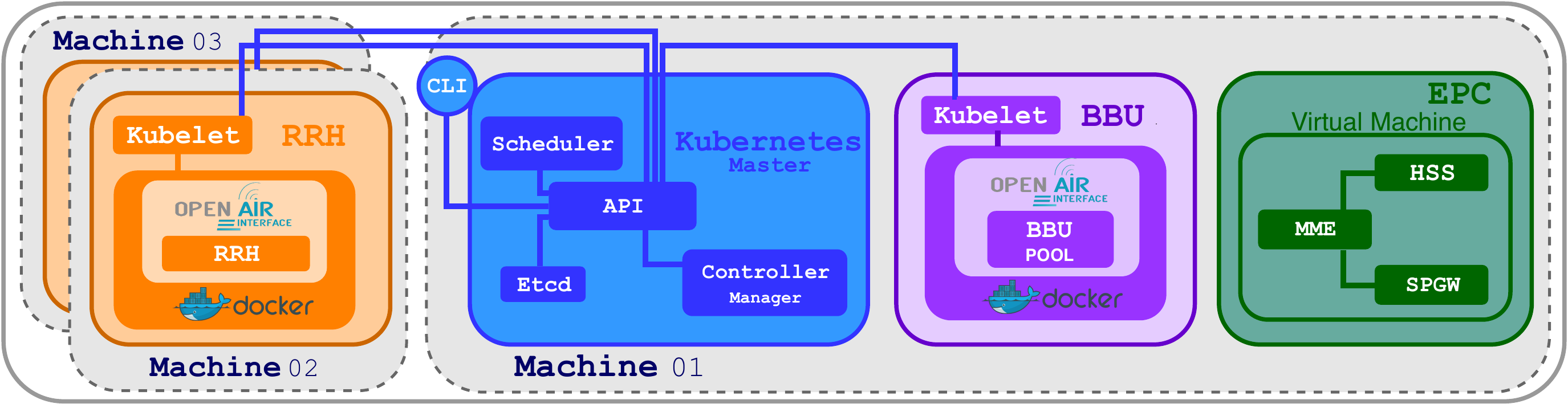}
    \caption{Orchestration system architecture with the main components of each module including
    the organization by machine used in the testbed.\label{fig:kube-system}}
\end{center}
\end{figure*}

Operating system-level virtualization, also known as containerization, refers to a feature that
enables the creation of multiple isolated user-spaces on top of a single operating system (OS)
kernel. A simplified way to create, deploy and manage containers is by using
Docker~\cite{docker}, the most popular containerization platform. This technology can be used
for telecommunications infrastructure, to increase density, scaling, speed of deployment,
portability and decrease cost~\cite{metaswitch_virtualization_nodate}. However, containers
also introduce new challenges and more complexity by creating an entirely new infrastructure.

In this context, orchestration platforms were designed to manage virtualized deployments in
large-scale clusters, alleviating the difficulty of maintaining a large number of standalone
containers. Orchestrators must address several important challenges including scalability,
fault tolerance, availability, efficient resource utilization, among
others~\cite{container-based-orchestration}. In this work, we adopted
Kubernetes~\cite{kubernetes}, which is the world's most popular open-source orchestration
system for Docker containers.

A Kubernetes cluster consists of at least one master and multiple computer nodes. The master
is responsible for exposing the application program interface (API), scheduling the deployments
and managing the overall cluster. The smallest unit in Kubernetes is a Pod, which consists of
one or more containers that share the same context and resources. Also, each node runs a
container runtime, such as Docker, along with kubelet, a Kubernetes agent for maintaining
the local pods according to the information provided by Kubernetes API.

\section{System Description and Results}

Figure~\ref{fig:kube-system} shows the developed system architecture, where RRH and BBU modules
of OAI were deployed on Docker containers. Dockerfiles were used to create Docker images for
BBU and RRU, with requirements based on OAI documentation~\cite{openairinterface_docker}.
For the EPC, all modules were compiled on a virtual machine since it will not be orchestrated.


BBU and RRH units for Kubernetes were made based on created Docker images. In order to deploy
theses units, StatefulSets were used. It's a Kubernetes controller used to scale a set of pods
and to guarantee ordering and uniqueness of them. To establish host and pod communication,
Calico~\cite{project-calico}, a Container Network Interface (CNI), was used to provide a layer
3 networking capability and associate a virtual router with each node.


In order to connect RRH with BBU, OAI requires some parameters: BBU, RRH, and EPC IP addresses
and also SIM card information when OAISIM is used. So, to be able to dynamically scale the BBU
and RRH, all needed information have to be stored on the ETCD, which is a distributed key-value
store. When BBU or RRH run, it accesses the ETCD using the container name as a key taking all
the values associated and storing its IP address. RRH run first and stores its IP address on
ETCD, after, BBU can run and take this information.





The system represented in Figure~\ref{fig:kube-system} was created on a local cluster with
three machines with 7th generation Intel(R) Core(TM) i5 processor and 8GB of RAM. One machine
acts as Kubernetes master and holds all the BBU needed pods and a virtual machine for EPC, as
depicted in Figure~\ref{fig:kube-system}. The other two are worker nodes and will hold one RRH
container each.

\begin{figure}[!h]
\begin{center}
\includegraphics[trim={1.2cm 2.5cm 1cm 1.8cm},clip,width=0.4\textwidth]{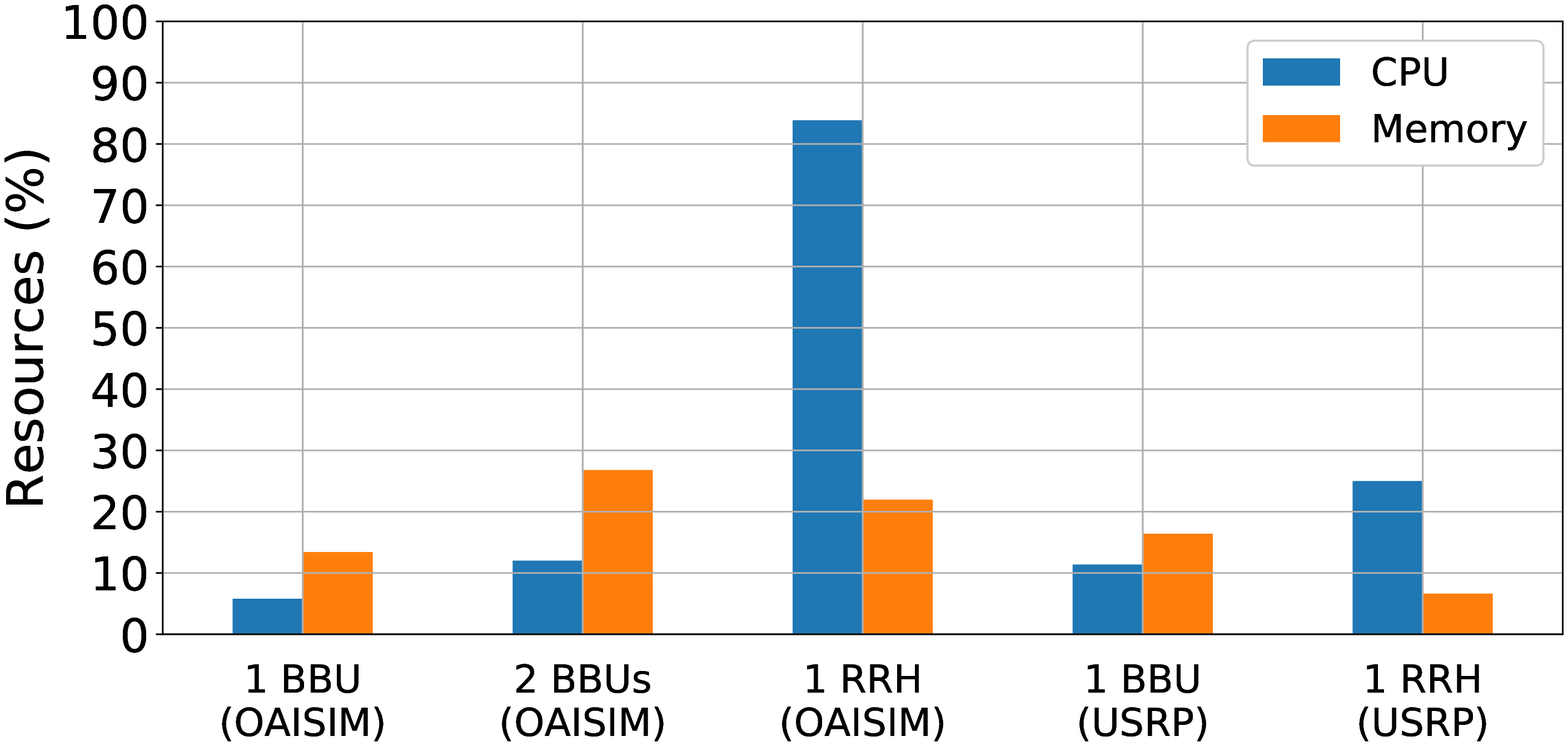}
\caption{Host's CPU and memory usage comparison of BBU and RRH with OAISIM and a real UE (USRP).\label{fig:resources}}
\end{center}
\end{figure}

Figure~\ref{fig:resources} shows the percentage of the host's resources used to run the BBU
and RRH. CPU and memory information can be used to create a load balancing policy, which could
allow dynamic changes to support network requirements. Also, the resources used by BBUs grow
proportionally, as can be seen by comparing the first two blocks. It also shows the OAISIM
requirement: more than 60\% of CPU usage when compared with the use of a real UE, since all
UE baseband processing is executed within RRH unit in the machine CPU.

\begin{figure}[!ht]
\begin{center}
\includegraphics[trim={3.4cm 0.5cm 1.2cm 2.9cm},clip,width=0.43\textwidth]{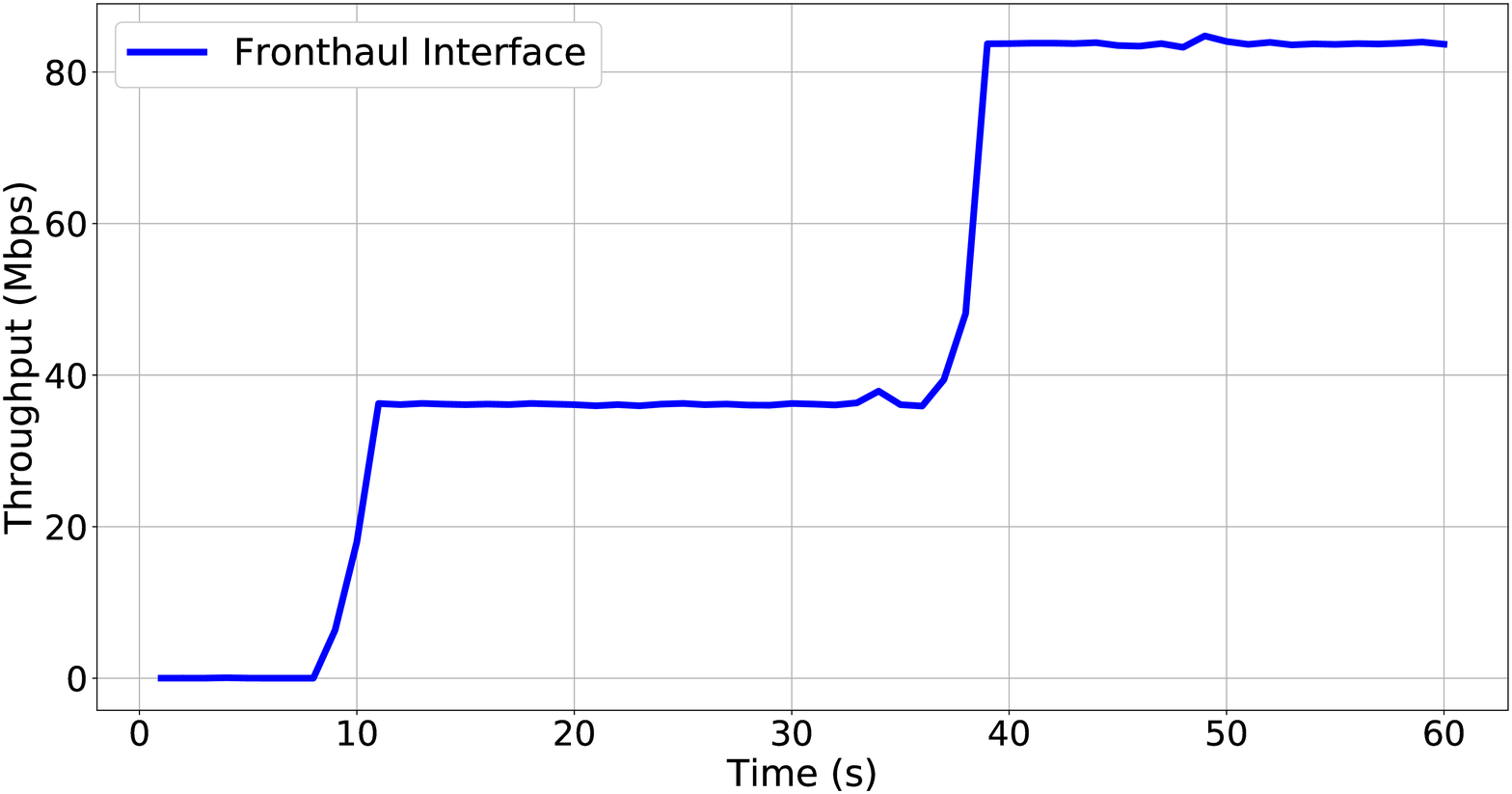}
\caption{Fronthaul throughput changing when a new instance of BBU and RRH starts its operation.\label{fig:fh}}
\end{center}
\end{figure}

Figure~\ref{fig:fh} shows the fronthaul throughput changing according to the number of BBU-RRU
pairs that are running. When the first BBU starts its operation, the throughput increases, and
then it stays constant. When the second BBU is scaled up, the impact can be observed in the
data rate, which doubles its value. Thus, this is a flexible and scalable system as we can
increase the number of BBU-RRU pairs in an easy way based on some metric or specifying a number.


\section{Conclusions}
This paper presented an orchestration of C-RAN using OAI with the most popular virtualization
technologies. This is a flexible system that can be used for research and development in any
cloud or local computing infrastructure, being possible to deploy LTE RAN stack elements in
order to allocate resources according to the aimed network scenario.


\bibliographystyle{IEEEtran}
\bibliography{references}

\end{document}